\documentclass[aps,prl,reprint,superscriptaddress,longbibliography,floatfix]{revtex4-2}
\usepackage{amsmath,amssymb,mathrsfs,bm,graphicx,hyperref,booktabs,xcolor}
\hypersetup{colorlinks=true,citecolor=blue,linkcolor=blue,urlcolor=blue}
\newcommand{\Vc}{\mathcal V}
\newcommand{\Ac}{\mathcal A}

\begin{document}

\title{Time Reflection-Refraction Crossover at Topological Kramers Edge States}
\author{Junpeng Hou}
\email{ryanhou94@gmail.com}
\affiliation{Pinterest Inc., San Francisco, California 94103, USA}
\affiliation{Department of Physics, Washington University in St. Louis, St. Louis, Missouri 63130, USA}
\author{Ya-Jie Wu}
\affiliation{School of Science, Xi'an Technological University, Xi'an 710032, China}

\begin{abstract}
Time-reversal symmetry protects helical edge states against static nonmagnetic backscattering. We show that spatially uniform temporal modulation produces a time reflection-refraction crossover within the same quantum spin Hall phase while preserving time-reversal symmetry of the instantaneous Hamiltonian. A finite gap to the bulk and an edge splitting that vanishes at the Kramers crossing allow the modulation to suppress bulk excitation while remaining nonadiabatic between the edge channels. Virtual transitions through bulk states enable coherent conversion between counter-propagating edge channels at fixed crystal momentum. Near the crossing, $qT$ governs the leading crossover dynamics for a fixed ramp shape, where $q$ is the momentum offset and $T$ the ramp duration. Coherent interference can suppress and restore reflection as the ramp slows. In the adiabatic limit, two initially time-reversed wave packets follow their respective propagation branches and acquire equal geometric displacements. Their group-velocity contributions cancel, isolating the contribution from the Berry curvature to be directly observable. Temporal modulation thus controls protected states and quantum geometry along a topological edge.
\end{abstract}

\maketitle

\emph{\textcolor{blue}{Introduction}}.
Reversing wave propagation need not require reversing momentum. A static spatial perturbation conserves energy, whereas a spatially uniform temporal modulation conserves momentum and can transfer amplitude between states of different energy and group velocity~\cite{Morgenthaler1958,Felsen1970,Fante1971,Mendonca2002,Mendonca2003,Galiffi2022}. This mechanism underlies temporal reflection and quantum time mirrors~\cite{Bacot2016,Reck2017,PachecoPena2020}, observed in electromagnetic, microwave, matter-wave, and elastic systems~\cite{Moussa2023,Dong2024,Jones2024,WangElastic2025}. Temporal scattering also provides access to band topology~\cite{Lustig2018,Wu2024,Xu2025,Xiong2025,Li2026,Jiang2026}. Recent work has further explored the effects of smooth temporal modulation and the displacement of scattered wave packets~\cite{Qin2024,Antyufeyeva2025,Deng2026}.

On a quantum spin Hall (QSH) edge, a static nonmagnetic perturbation cannot elastically couple a state at $k$ to its Kramers partner at $-k$, a consequence of protection by time-reversal symmetry (TRS)~\cite{KaneMele2005,KaneMeleZ2,BHZ2006,Konig2007,HasanKane2010,QiZhang2011,Ortiz2016,Vayrynen2018}. The corresponding temporal process connects channels at the same $k$, with the reversal encoded in their group velocities. Nonequilibrium helical-edge dynamics have been explored through quenches, electrical excitation, driven Rashba coupling, and wave-packet transport~\cite{Patel2013,Dolcini2016,Privitera2020,ZhangFoster2021,RazoLopez2025}. The distinction between the two scattering processes raises a basic question: can temporal modulation produce time reflection of a protected edge when every instantaneous Hamiltonian keeps TRS and the bulk remains in the same QSH phase? If so, what mediates the conversion, and how does it evolve as the time scale of the modulation changes?

We find in a Kane-Mele ribbon that bulk transitions can be negligible even when conversion between the edge channels remains appreciable. Direct projection of a TR-even control onto the degenerate doublet yields only a multiple of the identity. The conversion arises because the control changes the edge states through virtual transitions to the bulk. This changes the projected velocity matrix and rotates its eigenvectors, even though the control is scalar within a fixed crossing doublet.

Changing the ramp duration reveals coherent evolution within the edge doublet. Near the crossing, the reflection probability has a leading scaling form in $qT$, where $q$ is the momentum offset from the Kramers crossing and $T$ the ramp duration. This scaling applies when the ramp duration is varied while the ramp shape is held fixed. Amplitude transferred between branches at different stages accumulates a relative phase, allowing reflection to decrease and then grow again. For sufficiently slow modulation, each propagation channel follows its instantaneous eigenstate and time refraction emerges. TRS gives the partner branches opposite group velocities but equal mixed Berry curvature. Averaging the displacements of an initially TR-related pair of wave packets therefore cancels out ordinary propagation but retains the common geometric shift to be observed.

\begin{figure}[t]
\centering
\includegraphics[width=\columnwidth]{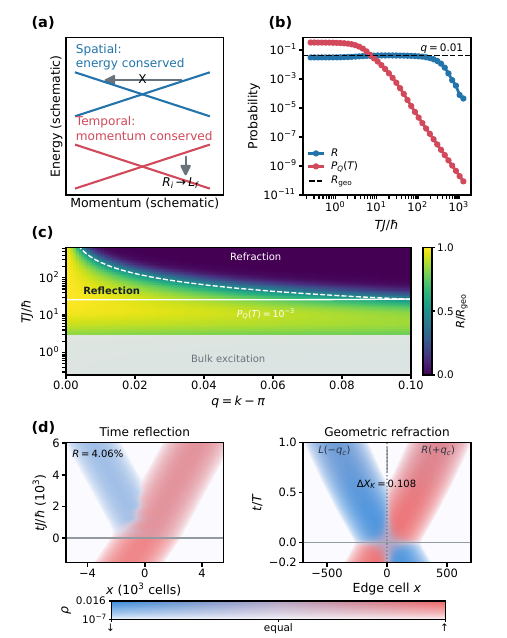}
\caption{\textbf{Temporal scattering at a Kramers edge.}
\textbf{(a)} Static backscattering would connect opposite momenta at fixed energy, a process forbidden by TRS. Time reflection connects opposite-velocity channels at fixed momentum.
\textbf{(b)} At $q=0.01$, final weight outside the edge doublet $P_Q(T)$ becomes small while reflection remains finite.
\textbf{(c)} Time reflection map in $(q,T)$. The gray band marks $T<\Delta_b^{-1}$. Solid and dashed white contours denote $P_Q(T)=10^{-3}$ and $R/R_{\rm geo}=1/2$, where $R_{\rm geo}$ is the limiting reflection probability with negligible bulk excitation.
\textbf{(d)} Examples of time reflective and refractive packets. Hue denotes spin polarization and brightness denotes density. All panels use the same control path: $\boldsymbol\lambda_i/J=(0.35,-0.50)$, $\boldsymbol\lambda_f/J=(-0.35,0)$, and $(\lambda_{\rm SO},\lambda_R)/J=(0.15,0.30)$, with $\boldsymbol\lambda=(m,g_y)$.}
\label{fig:one}
\end{figure}

\emph{\textcolor{blue}{Temporal scattering at a Kramers edge}}.
Given the TRS operator $\Theta$ and a Hermitian TR-even perturbation $V$, $\langle\Theta\psi|V|\psi\rangle=0$. Thus a static perturbation cannot elastically convert the right-moving state $R(k)$ into the left-moving one $L(-k)=\Theta R(k)$. A spatially uniform temporal modulation conserves $k$ instead of energy. It can therefore connect the initial $R_i(k)$ and final $L_f(k)$ with opposite velocities, as illustrated in Fig.~\ref{fig:one}(a).

This opens the possibility of time reflection of QSH edges, defined as $R_i(k)\to L_f(k)$, with group velocity satisfying $v_{R_i}(k)v_{L_f}(k)<0$. Time refraction carries the incident amplitude into the final branch with the same velocity sign. Their probabilities are $R(k)=|u_{f,L}^\dagger(k)\psi_k(T)|^2$ and $\mathscr T(k)=|u_{f,R}^\dagger(k)\psi_k(T)|^2$, where $\psi_k(T)$ is the normalized evolution state and $u_{f,R/L}(k)$ are normalized final right/left-moving modes. The remaining probability $P_Q$ is the weight outside the selected same-edge doublet including bulk and opposite edges. It is $P_Q(k,T)=\|[I-P_{f,e}(k)]\psi_k(T)\|^2$, with $P_{f,e}(k)=\sum_{\alpha=R,L}u_{f,\alpha}(k)u_{f,\alpha}^\dagger(k)$ the projector onto those two modes. Thus $R+\mathscr T+P_Q=1$.

We use a zigzag Kane-Mele ribbon with nearest-neighbor hopping $J$, intrinsic spin-orbit coupling $\lambda_{\rm SO}$, and Rashba coupling $\lambda_R$. We set $J=\hbar=1$ unless stated otherwise. As shown later, modulating the staggered sublattice potential $m$ alone produces time reflection and its crossover to refraction. We further introduce a second TRS control that changes the intracell spin-dependent hopping by $ig_ys_y$, where $s_y$ is the spin Pauli matrix. This control allows a nonzero geometric displacement and a path with near-total reflection. We denote $\boldsymbol\lambda=(m,g_y)$ with $m$ the mass. We primarily consider the controls following $\boldsymbol\lambda(t)=\boldsymbol\lambda_i+(\boldsymbol\lambda_f-\boldsymbol\lambda_i)f(s)$, where $s=t/T$, $T$ is the ramp duration, and the ramp shape is specified by $f(s)=[1-\cos(\pi s)]/2$. Supplemental Material (SM) gives full protocols and verifies that the bulk gaps remain open throughout the ramp.

Two spectral separations organize the response to this ramp. The gap $\Delta_b$ from the edge crossing to bulk states stays finite while the edge splitting $\Delta_e(q)$ vanishes as $q=k-k_0\to0$, with $k_0=\pi$. Adiabatic evolution with respect to the bulk can therefore coexist with nonadiabatic evolution between the edge channels, as confirmed in Fig.~\ref{fig:one}(b). As $T$ increases, the leakage $P_Q$ falls while reflection $R$ remains near a finite plateau, giving rise to time reflection. Figure~\ref{fig:one}(c) extends this comparison over momentum and duration, showing both reflection and refraction regions. The reflective interval reaches longer durations near the crossing because the edge splitting is smaller.

Figure~\ref{fig:one}(d) illustrates the propagation outcomes in these two regions using the same control path. For the reflective packet, part of the incident amplitude emerges on the negative-velocity branch and travels backward at the same momentum. For the refractive pair, the two initially TR-related packets follow their respective branches and move apart.

\begin{figure}[t]
\centering
\includegraphics[width=\columnwidth]{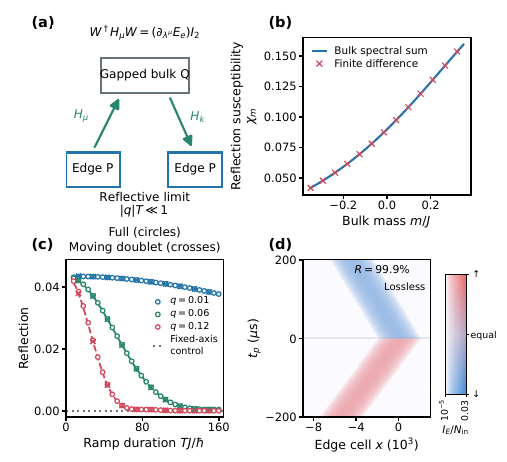}
\caption{\textbf{Bulk contribution to edge-channel conversion.}
\textbf{(a)} A TR-even control is scalar in a fixed crossing doublet. Virtual transitions through $Q$ change its velocity matrix.
\textbf{(b)} Mass reflection susceptibility $\chi_m$ from the bulk spectral sum and a parallel-transported finite difference.
\textbf{(c)} Full-ribbon (circles) and projected edge (crosses) dynamics. The mass ramp uses $m/J:-0.35\to0.35$, $g_y=0$, and $(\lambda_{\rm SO},\lambda_R)/J=(0.15,0.30)$. The fixed-axis control holds $\Vc$ and the initial and final velocity axes constant, giving $R=0$.
\textbf{(d)} Near-total time reflection, $R=99.9\%$, in the lossless coupled-mode model. Here $(\lambda_{\rm SO},\lambda_R)/J=(0.50,0.35)$ and $T=80$. The curved path runs from $\boldsymbol\lambda_i/J\simeq(-0.52,0)$ to $\boldsymbol\lambda_f/J\simeq(0.52,0)$.}
\label{fig:two}
\end{figure}

\emph{\textcolor{blue}{Virtual transitions and the reflective limit}}.
Let the columns of $W(\boldsymbol\lambda)$ form an orthonormal basis of the degenerate edge states, with $E_e$ the crossing energy $HW=E_eW$. Then $P=WW^\dagger$ projects onto this edge doublet, and $Q=I-P$ projects onto states outside it. For each control derivative $H_\mu=\partial_{\lambda^\mu}H$, TRS gives $W^\dagger H_\mu W=(\partial_{\lambda^\mu}E_e)I_2$. Thus, a direct control projection changes only the common energy.

The propagation direction is set by the projected velocity matrix $\Vc=W^\dagger H_kW$ with $H_k=\partial_kH|_{k_0}$. Although the control has no off-diagonal matrix element in a fixed crossing doublet, it changes the edge states themselves and hence $\Vc$. To describe this moving basis, define its Berry connection $\Ac_\mu=iW^\dagger\partial_{\lambda^\mu}W$~\cite{WilczekZee1984}. With negligible $P_Q$, assume the state near the crossing is $\psi\simeq Wc$, then the two edge amplitudes $c$ obey
\begin{equation}
i\dot c=\left[E_eI_2+q\Vc-\dot\lambda^\mu\Ac_\mu+O(q^2)\right]c,
\label{eq:edge}
\end{equation}
where repeated control indices are summed. The term $q\Vc$ resolves the channel splitting, while the connection accounts for the change of basis along the path.

A basis change within the doublet can rotate the matrix $\Vc$ without changing the physical channels. The velocity matrices must be compared in a transported basis. The covariant derivative $D_\mu\Vc=\partial_{\lambda^\mu}\Vc-i[\Ac_\mu,\Vc]$ removes this arbitrary basis rotation and gives
\begin{align}
D_\mu\Vc&=W^\dagger H_{k\mu}W+X_\mu+X_\mu^\dagger,\nonumber\\
X_\mu&=W^\dagger H_kQ(E_e-QHQ)^{-1}QH_\mu W .
\label{eq:response}
\end{align}
Here $H_{k\mu}=\partial_{\lambda^\mu}H_k$, and the inverse acts on the gapped bulk states of the isolated edge. The factor $QH_\mu W$ connects the edge to the bulk, and $H_k$ couples these states back to the edge velocity matrix. All controls considered here have $H_{k\mu}=0$, so their velocity response comes entirely from these virtual transitions, as sketched in Fig.~\ref{fig:two}(a). This changes the propagation basis without requiring a finite population of real bulk excitations in the slow limit (see SM for details).

For a small mass change $\delta m$, the resulting change of velocity eigenvectors gives the limiting reflection probability $R_{\rm geo}=\chi_m(\delta m)^2+O((\delta m)^3)$ with susceptibility $\chi_m$. Figure~\ref{fig:two}(b) compares $\chi_m$ obtained from the bulk spectral sum with a finite difference of the parallel-transported velocity eigenvectors and they match well. Figure~\ref{fig:two}(c) compares full-ribbon and projected edge evolution along the mass-only path. The agreement shows that the changing doublet captures the conversion. An analytic control with a fixed velocity matrix and bases gives no conversion despite spin mixing (see SM for details). Thus spin mixing alone does not produce time reflection.

The reflective plateau follows when the edge subspace tracks the control but the two propagation branches cannot be individually resolved. For a fixed smooth path, this corresponds to the interval $\Delta_b^{-1}\ll T\ll\Delta_e(q)^{-1}$. Choose a parallel frame with $W^\dagger\partial_tW=0$ and then the Berry connection vanishes. Write the traceless velocity matrix as $\Vc_{\rm tr}=\bm v\cdot\bm\sigma$, with doublet Pauli matrices $\bm\sigma$ and unit axis $\hat{\bm v}=\bm v/|\bm v|$. Its eigenvalues $\pm|\bm v|$ give the two velocity directions, and $\Delta_e\simeq2|q||\bm v|$.

In the joint limit $T\to\infty$, $q\to0$, and $|q|T\to0$, transitions to the bulk vanish while the evolution generated by $q\Vc_{\rm tr}$ becomes negligible. The amplitudes therefore retain their initial values in the parallel frame. The initial state is carried within the edge subspace and projecting it onto the final opposite-velocity channel gives
\begin{equation}
    2R_{\rm geo}=1-\hat{\bm v}_i\cdot\hat{\bm v}_f. \label{eq:Rgeo}
\end{equation}
Thus, two identical velocity axes give no reflection, while opposite axes give $R_{\rm geo}=1$. A quench generally excites bulk states and does not realize this limit. We choose a control path that realizes $R\simeq99.9\%$ in Fig.~\ref{fig:two}(d) in a coupled-mode model and details are given in SM.

\begin{figure}[t]
\centering
\includegraphics[width=\columnwidth]{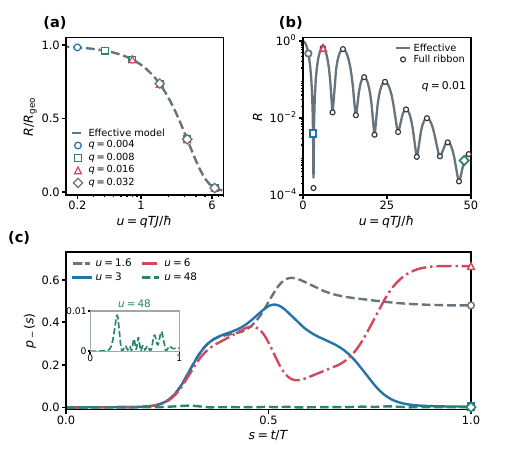}
\caption{\textbf{Reflection-refraction crossover.}
\textbf{(a)} Normalized reflection $R(q,T)/R_{\rm geo}$ on full ribbons with the effective prediction $\mathcal R(u)$ using parameters in Fig.~\ref{fig:one}(d).
\textbf{(b)} Reflection $R$ at $q=0.01$ using parameters from Fig.~\ref{fig:two}(d). Colored symbols select the durations in (c).
\textbf{(c)} Effective negative-velocity occupation $p_-(s)$ at different $u$ on the same path as (b). The inset enlarges $u=48$.}
\label{fig:three}
\end{figure}

\emph{\textcolor{blue}{Resolving the edge splitting and time reflection-refraction crossover}}.
The reflective plateau leaves open how reflection disappears at finite $q$ as shown in Fig.~\ref{fig:two}(c). For a fixed path, increasing $qT$ resolves the edge splitting and drives the crossover to refraction. In the parallel frame, removal of the common energy phase $\exp[-i\int_0^t E_e(\boldsymbol\lambda(t'))\,dt'/\hbar]$ gives
\begin{equation}
i\partial_s c=(qT)\Vc_{\rm tr}(s)c+O(q^2T),
\label{eq:qT}
\end{equation}
where the scaled duration $u=qT$ sets the phase accumulated between the branches. For a fixed control $\boldsymbol\lambda(s)$, the leading evolution depends only on $u$. Taking $q\to0$ at fixed nonzero $u$ sends $T\to\infty$, suppressing bulk excitation while the correction $q^2T$ vanishes. Thus $R(q,T)/R_{\rm geo}\to\mathcal R(u)$ as shown in Fig.~\ref{fig:three}(a). Here $\mathcal R(u)$ is the final negative-velocity probability from the leading part of Eq.~\eqref{eq:qT}, divided by $R_{\rm geo}\ne0$, for an initial positive-velocity channel.

The shape of this curve follows from the rotation of the velocity eigenvectors and their phase difference. Let $|v_\pm(s)\rangle$ have velocity eigenvalues $\pm v(s)$, with $v(s)=|\bm v(s)|>0$, and choose their phases so that $\langle v_\pm|\partial_s v_\pm\rangle=0$. Their rotation gives the nonadiabatic coupling $C(s)=\langle v_-(s)|\partial_s v_+(s)\rangle$. The edge splitting produces a relative dynamical phase $u\Phi(s)$, where $\Phi(s)=2\int_0^s v(s')\,ds'$. Define the branch amplitudes $b_\pm$ by $c(s)=b_+(s)e^{-iu\Phi(s)/2}|v_+(s)\rangle
+b_-(s)e^{iu\Phi(s)/2}|v_-(s)\rangle$,
they obey
\begin{equation}
\partial_s b_-=-C(s)e^{-iu\Phi(s)}b_+,~\partial_s b_+=C^*(s)e^{iu\Phi(s)}b_-.
\label{eq:channels}
\end{equation}
The coupling generates conversion, while the relative phase determines how amplitudes transferred at different stages interfere.

When $|u|\ll1$ and bulk excitation is negligible, the state changes little in the parallel frame. Projection onto the rotated final velocity basis then gives the geometric plateau, so $\mathcal R(0)=1$ in the joint limit of Eq.~\eqref{eq:Rgeo}. At intermediate $u$, converted amplitudes can cancel or enhance as the duration changes. This interference is pronounced on the near-total-reflection path already used in Fig.~\ref{fig:two}(d). Figure~\ref{fig:three}(b) shows that the reflection falls to about $0.4\%$ at $T=300$, then recovers to $66\%$ at $T=600$.

Thus, a small final reflection does not necessarily imply adiabaticity. For an incident positive-velocity channel, define $p_-(s)=|\langle v_-(s)|c(s)\rangle|^2=|b_-(s)|^2$, with $b_+(0)=1$ and $b_-(0)=0$. Its final value gives the effective reflection probability and Fig.~\ref{fig:three}(c) compares four durations on the same path. At $u=1.6$, conversion leaves a substantial reflected component. At $u=3$, the negative-velocity occupation reaches about $48\%$ before returning to a small final value. Despite substantial occupation of the negative-velocity channel during the ramp, destructive interference suppresses the final reflection. The $u=6$ trace instead retains the large reflected weight. At $u=48$, the negative-velocity channel remains below $1\%$ throughout the ramp.

For sufficiently large $|u|$, the state follows an individual velocity eigenvector throughout the ramp with nonzero splitting. Reflection then tends to zero and the velocity sign is retained, giving time refraction. Each branch accumulates dynamical and geometric phases whose momentum dependence determines the packet displacement. 

\begin{figure}[t]
\centering
\includegraphics[width=\columnwidth]{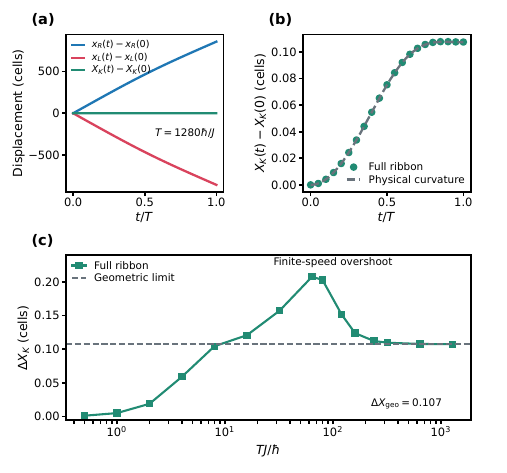}
\caption{\textbf{Geometric refraction of initially TR-related packets.}
\textbf{(a)} Full-ribbon physical displacements and their midpoint.
\textbf{(b)} Midpoint compared with the curvature integral averaged over the same incident spectrum.
\textbf{(c)} Midpoint displacement versus duration at fixed spectrum. The dashed line marks the slow ramp limit $\Delta X_{\rm geo}=0.107$ cells.}
\label{fig:four}
\end{figure}

\emph{\textcolor{blue}{Geometric displacement in the time refractive limit}}.
Once in the adiabatic limit, the changing eigenvector contributes a geometric velocity. For an instantaneous eigenstate $|u_n(k,\boldsymbol\lambda)\rangle$ of energy $E_n$, the physical packet center obeys~\cite{SundaramNiu1999,Xiao2010,Bleu2018Dynamics,Bleu2018QGT,Deng2026}
\begin{equation}
\dot x_n=\partial_kE_n-\dot\lambda^\mu\Omega^{(n)}_{k\mu},
\label{eq:semiclassical}
\end{equation}
where the first term is group velocity and the Berry curvature $\Omega^{(n)}_{k\mu}$ couples momentum to control $\lambda^\mu$. Its velocity scales with the ramp rate, but its integrated displacement depends on the control path in the adiabatic limit.

The physical position includes the location of each orbital within a longitudinal cell. We can define the cell-periodic curvature $\Omega^{(n),\rm cell}_{k\mu}=-2\operatorname{Im}\langle\partial_ku_n|\partial_{\lambda^\mu}u_n\rangle$ and the intracell-coordinate operator $\hat X_\parallel$. The physical curvature is then ${\Omega^{(n)}_{k\mu}=\Omega^{(n),\rm cell}_{k\mu}-\partial_{\lambda^\mu}\langle u_n|\hat X_\parallel|u_n\rangle}$. The second term accounts for the change of intracell position as the eigenstate evolves. Their sum, and hence the physical displacement, is independent of the assignment of orbitals to unit cells (see SM). On the mass-only path, the two contributions cancel while the two-control path gives a nonzero physical longitudinal geometric displacement, highlighting the time-refractive response.

This geometric displacement can be much smaller than the distance traveled at the group velocity and TRS provides a way to remove that background. The partner branches $|u_L(-k)\rangle=\Theta|u_R(k)\rangle$ have opposite group velocities but equal mixed curvature. So the group-velocity contributions cancel in their midpoint $X_K=(x_R+x_L)/2$, while the two geometric contributions add. Integrating Eq.~\eqref{eq:semiclassical} then gives
\begin{equation}
\Delta X_K=-\int_{\boldsymbol\lambda_i}^{\boldsymbol\lambda_f}
d\lambda^\mu \left(\sum_jw_j\Omega^R_{k\mu}(k_j,\boldsymbol\lambda)\right),
\label{eq:drag}
\end{equation}
where the momentum probabilities satisfy $\sum_jw_j=1$ and they are conserved during the adiabatic process.

Figure~\ref{fig:four}(a) illustrates the separation in time refraction, where each packet travels hundreds of cells and the midpoint shifts by about $0.11$ cells. Figure~\ref{fig:four}(b) compares this midpoint throughout the ramp with the curvature integral averaged over the same incident spectrum and the two curves follow one another. Figure~\ref{fig:four}(c) varies the ramp duration at fixed path and incident spectrum. At intermediate durations, the midpoint overshoots the adiabatic prediction because the exact states do not yet follow individual branches perfectly. As $T$ increases, it eventually converges.

\emph{\textcolor{blue}{Discussion}}.
Temporal modulation separates confinement to a topological edge from adiabatic following of its propagation channels. The finite bulk gap allows the edge subspace to track the control while the vanishing splitting near the Kramers crossing leaves its two branches unresolved. In this interval, virtual transitions through bulk states change the velocity eigenvectors, and their mismatch after transport produces time reflection. The crossover is organized by $qT$ for a fixed control path and ramp shape. Relative phases acquired during channel conversion allow reflection to decrease and recover before the evolution approaches adiabatic time refraction.

The refractive limit retains information beyond the channel probabilities. TR-related branches acquire equal geometric displacements, which can be measured by averaging two packets with matched spectra and is independent of ramp duration in the adiabatic limit. Temporal modulation therefore controls the evolution of the propagation basis through reflection and the geometry of a followed branch through displacement. Extensions to larger spin systems, including spin-tensor Hall insulators~\cite{Wu2025QSTH}, could examine topological tensor responses at the temporal interface. Our work reveals temporal modulation as a meaningful tool to control and manipulate a spinful and topological edge.

\emph{\textcolor{blue}{Acknowledgments}}
J. Hou acknowledges the helpful discussions and hospitality of C. Zhang during the visit at WUSTL.  
Y.-J. Wu was supported by NSFC under the grant No.12275203, Shaanxi Fundamental Science Research Project for Mathematics and Physics (Grant No.25JSY002).
\bibliography{references}

\clearpage
\onecolumngrid
\setcounter{section}{0}
\setcounter{subsection}{0}
\setcounter{secnumdepth}{3}
\renewcommand{\thesection}{S\arabic{section}}
\renewcommand{\theHsection}{sm.\arabic{section}}
\begin{center}\large\textbf{Supplemental Material for ``Time Reflection-Refraction Crossover at Topological Kramers Edge States''}\end{center}
\setcounter{equation}{0}
\renewcommand{\theequation}{S\arabic{equation}}
\renewcommand{\theHequation}{sm.\arabic{equation}}
\setcounter{figure}{0}\setcounter{table}{0}
\renewcommand{\thefigure}{S\arabic{figure}}
\renewcommand{\thetable}{S\arabic{table}}
\renewcommand{\theHfigure}{sm.\arabic{figure}}
\renewcommand{\theHtable}{sm.\arabic{table}}

\section{Hamiltonian, symmetry, and setup}
\label{app:scope}
We consider noninteracting single-particle dynamics under spatially uniform controls $\boldsymbol\lambda(t)$. We denote $k_0$ the Kramers crossing and the momentum offset $q=k-k_0$. The two-dimensional bulk momentum is $\bm k=(k_1,k_2)$. We use $J=\hbar=1$ unless units are shown explicitly. Times and longitudinal displacements are measured in $\hbar/J$ and number of cells. Every instantaneous Hamiltonian obeys time-reversal symmetry (TRS)
$\Theta H(\bm k,\boldsymbol\lambda)\Theta^{-1}=H(-\bm k,\boldsymbol\lambda)$, with $\Theta^2=-1$.
For the Kane-Mele model, $\Theta=is_yK$, with complex conjugation $K$ and spin Pauli matrices $s_\mu$. Pauli matrices $\sigma_\mu$ act within a projected edge doublet.

At each conserved $k$, the final same-edge eigenvectors $u_{f,R}(k)$ and $u_{f,L}(k)$ have positive and negative group velocity, respectively. Their projector is
$P_{f,e}(k)=\sum_{\alpha=R,L}u_{f,\alpha}(k)u_{f,\alpha}^\dagger(k)$.
For a normalized state $\psi_k(T)$ evolved from the positive-velocity input, define
\begin{align}
\mathscr T(k)&=|u_{f,R}^\dagger(k)\psi_k(T)|^2,\nonumber\\
R(k)&=|u_{f,L}^\dagger(k)\psi_k(T)|^2,\nonumber\\
P_Q(k,T)&=\|[I-P_{f,e}(k)]\psi_k(T)\|^2.
\label{app:eq:observables}
\end{align}
A packet probability is the sum over conserved momenta with its incident spectral weights and unitary evolution gives $R+\mathscr T+P_Q=1$. At intermediate times, $P_Q(k,t)$ is defined using the instantaneous edge projector.

A finite ribbon contains another edge doublet that can become nearly degenerate with the selected one. We resolve both edge sectors before identifying the remaining bulk modes. Their orthogonal projectors $\mathsf P_e$ and $\mathsf P_{\rm far}$ define $\mathsf P_b=I-\mathsf P_e-\mathsf P_{\rm far}$. The probabilities $P_b=\|\mathsf P_b\psi\|^2$ and $P_{\rm far}=\|\mathsf P_{\rm far}\psi\|^2$ then obey $P_Q=P_b+P_{\rm far}$. Accordingly, the bulk isolation gap excludes both edges.

\section{Kane-Mele model and control paths}
\label{app:km}
The Kane-Mele model includes intrinsic spin-orbit coupling~\cite{KaneMele2005,KaneMeleZ2} and Rashba spin mixing~\cite{Ortiz2016}. A zigzag ribbon has transverse rows indexed by $j$, with basis $(A\uparrow,A\downarrow,B\uparrow,B\downarrow)$. The mass matrix is $M=\operatorname{diag}(1,1,-1,-1,\ldots)$. The second control changes each intracell $A_j$ to $B_j$ hopping by $ig_ys_y$, with the reverse hopping given by its Hermitian conjugate. Denoting this control as $g_yG_y$, the Hamiltonian in the cell Bloch convention is
\begin{equation}
H(k,m,g_y)=H_{\rm KM}(k)+mM+g_yG_y,
\label{app:eq:bloch}
\end{equation}
where $H_{\rm KM}(k)$ is the unmodulated ribbon Hamiltonian. Both controls are TR even and they are independent of $k$, giving $H_{km}=H_{kg_y}=0$.

The mass-only path in Fig.~\ref{fig:two}(b,c) uses $g_y=0$, a mass ramp from $-0.35$ to $0.35$, and $(\lambda_{\rm SO},\lambda_R)=(0.15,0.30)$. At zero Rashba coupling, the two spin sectors are Haldane models with valley masses $m\pm3\sqrt3\lambda_{\rm SO}$ so the QSH interval is $|m|<3\sqrt3\lambda_{\rm SO}$. A sampled gapped interpolation from the finite-Rashba mass path to this spin-conserving limit has minimum direct bulk gap approximately $0.24J$. The crossing-to-bulk separation along the mass path is approximately $0.34J$. The path in Figs.~\ref{fig:one}, \ref{fig:three}(a), and \ref{fig:four} uses the same spin-orbit couplings and
$m(s)/J=0.35-0.70f(s)$ and $g_y(s)/J=-0.50+0.50f(s)$, with $f(s)=[1-\cos(\pi s)]/2$ and $s=t/T$.
Numerical results give a minimum direct bulk gap $0.24J$. The sampled crossing-to-bulk isolation scale is $\Delta_b\simeq0.34J$, with both edges excluded. The final point lies on the mass-only QSH path, so the open direct bulk gap preserves the QSH phase along the whole control path. The finite gap permits transport of the edge subspace with negligible bulk excitation.

\section{Scalar control projection and bulk response}
\label{app:response}
At $k_0$, write the single-edge frame as $W=(|e_1\rangle,|e_2\rangle)$, with $HW=E_eW$, $W^\dagger W=I_2$, $P=WW^\dagger$, and $Q=I-P$. In a Kramers basis, TR acts as $i\sigma_yK$. A Hermitian $2\times2$ matrix with the form $a_0I_2+\bm a\cdot\bm\sigma$ has $\bm a=0$ under TRS. Differentiating the degenerate eigenvalue equation then gives
$W^\dagger H_\mu W=\eta_\mu I_2$, where $H_\mu=\partial_{\lambda^\mu}H$ and $\eta_\mu=\partial_{\lambda^\mu}E_e$.

The velocity matrix and doublet Berry connection are
$\Vc=W^\dagger H_kW$, with $H_k=\partial_kH|_{k_0}$, and $\Ac_\mu=iW^\dagger\partial_{\lambda^\mu}W$.
For negligible bulk excitation, substituting $\psi\simeq Wc$ into the Schr\"odinger equation and expanding in $q$ yields
\begin{equation}
i\dot c=[E_eI_2+q\Vc-\dot\lambda^\mu\Ac_\mu+O(q^2)]c .
\label{app:eq:movingdoublet}
\end{equation}
The coefficient vector $c$ describes amplitudes in the changing edge basis. Under $W\to WU(\boldsymbol\lambda)$, with $U$ a $2\times2$ unitary basis change, $\Vc\to U^\dagger\Vc U$ and $\Ac_\mu\to U^\dagger\Ac_\mu U+iU^\dagger\partial_{\lambda^\mu}U$. Consequently,
$D_\mu\Vc=\partial_{\lambda^\mu}\Vc-i[\Ac_\mu,\Vc]$
transforms covariantly. A parallel frame along the path has $W^\dagger\partial_tW=0$ and it compares the initial and final channel axes in a common transported basis.

Projecting the derivative of $HW=E_eW$ onto $Q$ gives
\begin{equation}
Q\partial_{\lambda^\mu}W
=Q(E_e-QHQ)^{-1}QH_\mu W .
\end{equation}
The inverse is restricted to the gapped bulk subspace of the isolated edge. Differentiating $\Vc$ and removing the in-frame derivative gives
\begin{align}
D_\mu\Vc&=W^\dagger H_{k\mu}W+X_\mu+X_\mu^\dagger,
\label{app:eq:generalresponse}\\
X_\mu&=W^\dagger H_kQ(E_e-QHQ)^{-1}QH_\mu W\nonumber\\
&=\sum_{b\in Q}
\frac{W^\dagger H_k|b\rangle\langle b|H_\mu W}{E_e-E_b}.
\label{app:eq:spectralsum}
\end{align}
Here $H_{k\mu}=\partial_{\lambda^\mu}H_k$, and $|b\rangle$ has energy $E_b$. The energy denominators measure the separation between the edge doublet and each bulk state. The control couples the edge to the bulk through $H_\mu$, and $H_k$ couples this response back to the velocity matrix. In the Bloch convention of Eq.~\eqref{app:eq:bloch}, the direct derivative vanishes for both controls, so this virtual bulk response accounts for the change of the edge velocity matrix. Figure~\ref{fig:two}(b) compares the resulting susceptibility with the change of parallel-transported velocity eigenvectors.

\section{Reflection with negligible bulk excitation}
\label{app:selective}
Write the traceless velocity matrix as $\Vc_{\rm tr}=\bm v\cdot\bm\sigma$ and $\hat{\bm v}=\bm v/|\bm v|$. Its eigenvalues $\pm|\bm v|$ give the opposite propagation directions near $k_0$, with energy splitting $\Delta_e(q)\simeq2|q||\bm v|$. For sufficiently small $q$, bulk isolation can hold while the accumulated internal evolution is small. The schematic interval
$\Delta_b^{-1}\ll T\ll\Delta_e(q)^{-1}$
expresses this separation. Bulk adiabaticity is set by the bulk gap and control matrix elements, while the internal evolution is set by the accumulated edge splitting.

After removal of the scalar phase, we have $q\Vc_{\rm tr}+O(q^2)$ in a parallel frame. In the limit
$T\to\infty$, $q\to0$, and $|q|T\to0$,
the edge subspace follows the path, while $c$ stays fixed in that frame. The velocity projectors are
$\Pi_R=(I_2+\hat{\bm v}\cdot\bm\sigma)/2$ and $\Pi_L=(I_2-\hat{\bm v}\cdot\bm\sigma)/2$.
The output probability is therefore
\begin{equation}
R_{\rm geo}=\operatorname{tr}(\Pi_{L,f}\Pi_{R,i})
=\frac{1-\hat{\bm v}_i\cdot\hat{\bm v}_f}{2},
\label{app:eq:geo}
\end{equation}
where the bulk excitation vanishes and both velocity axes belong to the same parallel frame. A small mass change $\delta m=m_f-m_i$ gives $R_{\rm geo}=\chi_m(\delta m)^2+O((\delta m)^3)$, with
\begin{equation}
\chi_m=\frac14|D_m\hat{\bm v}|^2
=\frac{|D_m\bm v|^2-(\hat{\bm v}\cdot D_m\bm v)^2}{4|\bm v|^2}.
\label{app:eq:chi}
\end{equation}
Here $D_m\bm v$ is the coefficient vector of $D_m\Vc_{\rm tr}$, and $D_m\hat{\bm v}$ is its induced derivative on the unit axis. The susceptibility $\chi_m$ measures the local change of propagation direction per squared mass change. Only the response perpendicular to the velocity axis contributes.

\section{Crossover and edge scattering matrix}
\label{app:crossover}
In the parallel frame, removal of the scalar phase from Eq.~\eqref{app:eq:movingdoublet} gives $i\partial_s c=(qT)\Vc_{\rm tr}(s)c+O(q^2T)$, with $s=t/T$. The scaled duration is $u=qT$ in our units, or $qTJ/\hbar$ with units restored. For $R_{\rm geo}\ne0$, keeping the control functions $\boldsymbol\lambda(s)$ fixed gives $R(q,T)/R_{\rm geo}\to\mathcal R(u)$ as $q\to0$ at nonzero fixed $u$. The correction $q^2T=qu$ vanishes while the evolution becomes slow relative to the bulk. If $|\bm v(s)|$ remains nonzero, adiabaticity gives $\mathcal R(u)\to0$ as $|u|\to\infty$. The intermediate curve depends on both the control path and the ramp shape.

Let $|v_\pm(s)\rangle$ be the eigenvectors of $\Vc_{\rm tr}(s)$ with eigenvalues $\pm v(s)$, where $v(s)=|\bm v(s)|>0$, and choose $\langle v_\pm|\partial_s v_\pm\rangle=0$. Define $G(s)=\int_0^s v(s')\,ds'$. Expanding the state as
\begin{equation}
c(s)=b_+(s)e^{-iuG(s)}|v_+(s)\rangle+b_-(s)e^{iuG(s)}|v_-(s)\rangle
\label{app:eq:branchamplitudes}
\end{equation}
and projecting its derivative onto the two eigenvectors gives Eq.~\eqref{eq:channels}, with $C=\langle v_-|\partial_s v_+\rangle$ and $\Phi=2G$. Orthonormality gives $\langle v_+|\partial_s v_-\rangle=-C^*$ and ensures $\partial_s(|b_+|^2+|b_-|^2)=0$. The phase convention removes the diagonal branch connections from the evolution, while their effect remains in the phase of $C$. The local ratio $|C(s)|/[2|u|v(s)]$ compares the rate of basis rotation with the branch splitting. Interference can also produce a small final $|b_-|^2$ after substantial occupation of the negative-velocity channel during the ramp.

Let $W_i(k)$ and $W_f(k)$ contain the initial and final velocity eigenvectors, ordered as $R,L$, and let $U(k,T)$ be the full-ribbon propagator. The edge scattering matrix is $S_e=W_f^\dagger UW_i$, with columns labeling the input and rows the output. Thus $|(S_e)_{LR}|^2$ is the reflected probability for an $R$ input. For $Q_f=I-W_fW_f^\dagger$ and $B=Q_fUW_i$, unitarity gives $S_e^\dagger S_e+B^\dagger B=I_2$. The singular values of $S_e$ measure final loss from the edge doublet.

\begin{figure}[ht]
\centering
\includegraphics[width=.96\textwidth]{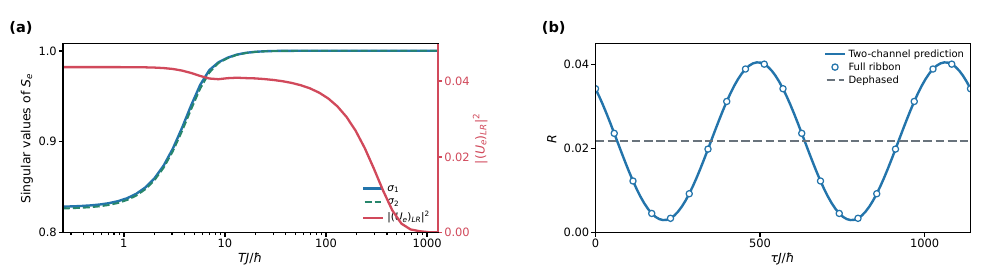}
\caption{Support for coherent edge conversion. \textbf{(a)} Singular values of $S_e$ (left axis) and mean off-diagonal probability of $U_e$ (right axis) at $q=0.01$. These measure final confinement and conversion, respectively. \textbf{(b)} Reflection at $q=0.008$ for two fixed ramps with a variable intermediate hold. The solid line is Eq.~\eqref{app:eq:holdinterference}, symbols are full-ribbon evolution, and the dashed line removes middle-channel coherences while retaining their populations.}
\label{app:fig:crossover_support}
\end{figure}

For nonsingular $S_e$, the unitary factor $U_e=S_e(S_e^\dagger S_e)^{-1/2}$ separates conversion from norm loss. The two off-diagonal probabilities of a $2\times2$ unitary matrix are equal. Figure~\ref{app:fig:crossover_support}(a) shows their mean together with both singular values at $q=0.01$. The singular values approach unity while conversion remains finite. When the branches are followed adiabatically, $S_e$ approaches $\operatorname{diag}(e^{i\phi_R},e^{i\phi_L})$, where $\phi_{R,L}$ contain dynamical and geometric phases.

To isolate the role of relative phase, split the control path at $(m,g_y)=(0,-0.25)$ into two raised-cosine ramps, each lasting $58.5\hbar/J$. Keep both ramps fixed and vary only the duration $\tau$ of a hold at this intermediate point. For $q=0.008$, let $a_\pm$ be the middle-channel amplitudes produced by the first ramp and $d_\pm$ the amplitudes for the second ramp to carry each middle channel into the final reflected channel. The middle-channel energies are $E_\pm$, with splitting $\Delta_e=E_+-E_->0$. When occupation outside the doublet is negligible, the reflected amplitude is
\begin{equation}
r(\tau)=d_-a_-e^{-iE_-\tau}+d_+a_+e^{-iE_+\tau}.
\label{app:eq:holdinterference}
\end{equation}
Its probability contains an interference term proportional to $\cos(\Delta_e\tau+\varphi_0)$, where $\varphi_0=\arg[d_-a_-(d_+a_+)^*]$ is their relative phase. All amplitudes and phases are determined by the fixed ramps. The oscillation period is $2\pi/\Delta_e$ and the full-ribbon splitting $\Delta_e\simeq0.011J$ predicts a period of about $570\hbar/J$. Figure~\ref{app:fig:crossover_support}(b) compares this two-channel prediction with full-ribbon evolution.

An ideal dephasing operation at the start of the hold removes the coherences between the two instantaneous channels while retaining their populations. In the full ribbon we apply $\rho\mapsto P_-\rho P_-+P_+\rho P_++Q\rho Q$, where $P_\pm$ project onto the middle edge eigenstates and $Q=I-P_- -P_+$. The density matrix $\rho$ retains unit trace and all three sector weights. The oscillations disappear when the same second ramp is applied. This diagnostic isolates the interference between edge-channel amplitudes.

\section{Geometric refraction and physical displacement}
\label{app:refraction}
\subsection{Physical velocity and mixed Berry curvature}
For a nondegenerate edge branch of energy $E_n$ followed adiabatically, the physical packet velocity is
$\dot x_n=\partial_kE_n-\dot\lambda^\mu\Omega^{(n)}_{k\mu}$.
The Bloch convention in Eq.~\eqref{app:eq:bloch} represents physical position by $i\partial_k+\hat X_\parallel$. The diagonal intracell-coordinate operator has entries $j/2$ on $A_j$ and $(j+1)/2$ on $B_j$. Its velocity vertex is
\begin{equation}
V_k^{\rm phys}=H_k-i[\hat X_\parallel,H].
\label{app:eq:physicalvelocity}
\end{equation}
The commutator has zero expectation in an eigenstate, so its group velocity is still $\partial_kE_n$ but it is essential in interband response. The physical mixed curvature is
\begin{equation}
\Omega^{(n)}_{k\mu}
=-2\operatorname{Im}\sum_{b\ne n}
\frac{\langle n|V_k^{\rm phys}|b\rangle
\langle b|H_\mu|n\rangle}{(E_n-E_b)^2},
\label{app:eq:mixedcurvature}
\end{equation}
where the sum includes all other branches, including the other same-edge channel. In a finite ribbon, far-edge modes are treated as the decoupled sector described in Sec.~\ref{app:scope}. Replacing $V_k^{\rm phys}$ by $H_k$ defines the cell-periodic curvature
$\Omega^{(n),\rm cell}_{k\mu}=-2\operatorname{Im}
\langle\partial_ku_n|\partial_{\lambda^\mu}u_n\rangle$.
Using $\langle n|[X,H]|b\rangle=(E_b-E_n)\langle n|X|b\rangle$ in the spectral sum gives
\begin{equation}
\Omega^{(n)}_{k\mu}
=\Omega^{(n),\rm cell}_{k\mu}
-\partial_{\lambda^\mu}\langle n|\hat X_\parallel|n\rangle,
\label{app:eq:embeddingcorrection}
\end{equation}
which fixes the sign for the connection $i\langle u|\partial u\rangle$. Changing which cell contains an orbital redistributes the two terms while preserving their sum. At the crossing, $W^\dagger[\hat X_\parallel,H]W=0$, so the projected physical velocity agrees with $\Vc$ in Sec.~\ref{app:response}. Away from the crossing, the curvature includes both bulk and edge interband contributions while Eq.~\eqref{app:eq:spectralsum} describes the virtual bulk contribution to the velocity response.

\subsection{TR pairing and packet midpoint}
For $|u_L(-k)\rangle=\Theta|u_R(k)\rangle$, the energy relation $E_L(-k)=E_R(k)$ gives opposite group velocities. The physical velocity vertex is TR odd and $H_\mu$ is TR even. Their product in Eq.~\eqref{app:eq:mixedcurvature} changes to minus its complex conjugate. Thus
$v_L(-k)=-v_R(k)$ and $\Omega^L_{k\mu}(-k)=\Omega^R_{k\mu}(k)$.
With normalized, time-reversed spectral weights $w_j$, define
$\overline\Omega^R_{k\mu}=\sum_jw_j\Omega^R_{k\mu}(k_j,\boldsymbol\lambda)$.
The midpoint $X_K=(x_R+x_L)/2$ of the two adiabatically followed packets then satisfies
$\dot X_K=-\dot\lambda^\mu\overline\Omega^R_{k\mu}$ and $\Delta X_K=-\int d\lambda^\mu\overline\Omega^R_{k\mu}$.
On the mass-only path at $g_y=0$, the cell-periodic curvature is canceled by the intracell derivative, giving zero physical longitudinal shift while the two-control path gives a nonzero physical longitudinal shift as shown in the main text.

\subsection{Exact displacement and the adiabatic limit}
The exact simulation reads displacement from the full evolving states through the physical current,
\begin{equation}
\Delta x_\alpha(t)=\sum_jw_j\int_0^t dt'\,
\langle\psi_{\alpha j}(t')|V_k^{\rm phys}(k_{\alpha j},t')|\psi_{\alpha j}(t')\rangle.
\label{app:eq:currentreadout}
\end{equation}
The pair midpoint is $\Delta X_K(t)=[\Delta x_R(t)+\Delta x_L(t)]/2$. Here $k_{Rj}=k_j$, $k_{Lj}=-k_j$ modulo $2\pi$, and each component starts in the corresponding TR-related eigenstate. Initial TR pairing does not imply that the two exact states remain TR partners under the same forward-time protocol. Accordingly, Eq.~\eqref{app:eq:currentreadout} includes finite-speed corrections, while Eq.~\eqref{eq:drag} is the adiabatic prediction.

\begin{figure}[t]
\centering
\includegraphics[width=.98\textwidth]{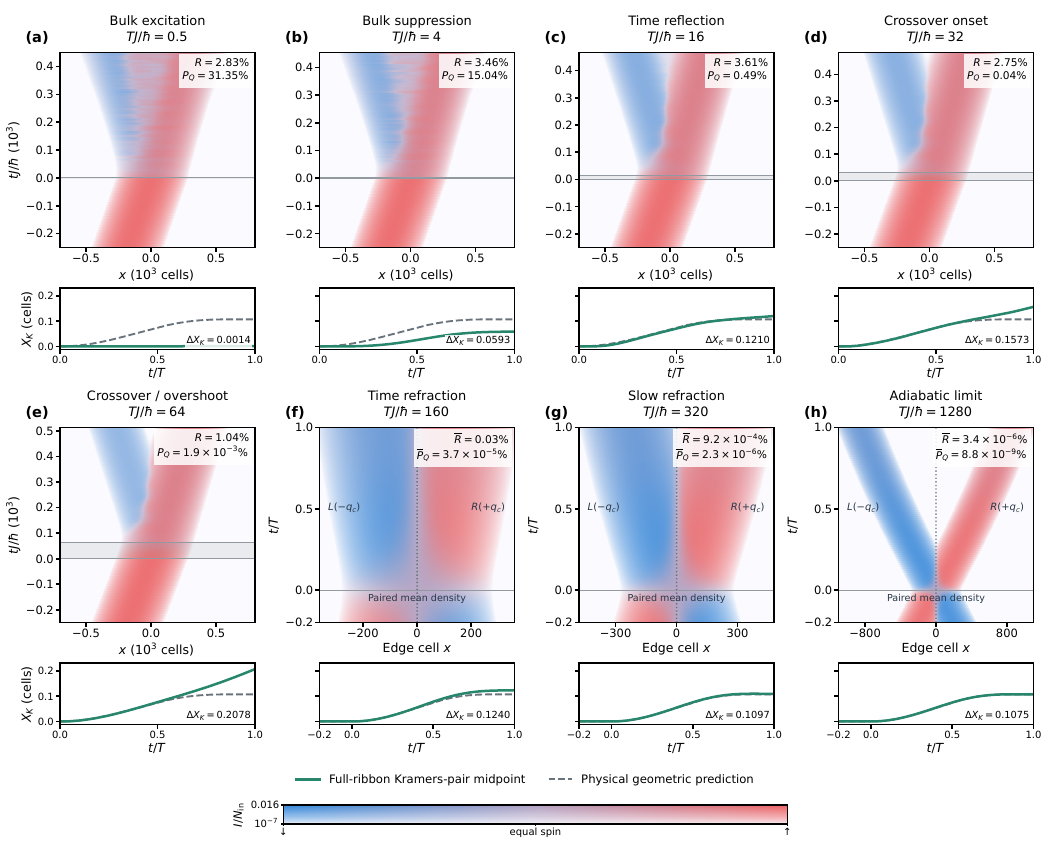}
\caption{Wave-packet evolution across ramp durations at fixed control path. The ramp duration increases from \textbf{(a)} to \textbf{(h)}. Upper panels \textbf{(a-e)} show the positive-velocity input before, during, and after the ramp. Panels \textbf{(f-h)} show the mean density of independently evolved $R(+q_c)$ and $L(-q_c)$ packets during the ramp, against $t/T$. Dotted lines mark their physical midpoint. The lower panels show pair midpoint displacement (green) and the spectrum-averaged adiabatic prediction (dashed gray), on a common displacement scale.}
\label{app:fig:regimes}
\end{figure}

Figure~\ref{app:fig:regimes} varies only $T$ at fixed control path and packet spectrum. The short ramp in Fig.~\ref{app:fig:regimes}(a) excites bulk states, whose weight decreases in Fig.~\ref{app:fig:regimes}(b). The backward packet becomes visible in Fig.~\ref{app:fig:regimes}(c). As $T$ increases further, Figs.~\ref{app:fig:regimes}(d,e) show decreasing reflection while the midpoint overshoots its geometric limit. Figures~\ref{app:fig:regimes}(f,g,h) show the mean density of two independently evolved TR-related inputs as their midpoint approaches that limit. All lower panels compare the physical pair midpoint with the same curvature prediction. This connects the changes in channel population to the convergence of the refractive displacement.

\section{A fixed-axis control with spin mixing}
\label{app:exact}
A conserved spin component fixes the propagation axis even with finite spin mixing. Consider the square-lattice Hamiltonian
\begin{equation}
H_c(p,k_y,m)=(m+\cos p+\cos k_y)\tau_z+\sin k_y\tau_y
+\sin p\,\tau_x(s_z+rs_x).
\label{app:eq:square}
\end{equation}
Here $\tau_\mu$ act on two orbitals, $s_\mu$ on spin, $p$ is momentum along the boundary, $k_y$ is transverse bulk momentum, and $r$ is a dimensionless spin-mixing strength. The Hamiltonian preserves TRS $\Theta=is_yK$ and its bulk spectrum is $E^2=(m+\cos p+\cos k_y)^2+\sin^2k_y+(1+r^2)\sin^2p$. At $r=0$, $-2<m<0$ is QSH. The nonnegative $r^2\sin^2p$ term preserves the bulk gap as $r$ increases.

The normalized spin operator $(s_z+rs_x)/\sqrt{1+r^2}$ commutes with the Hamiltonian for every $m$. At the lower boundary and $p=0$, the edge states have orbital component $\tau_x=-1$, giving the projected velocity matrix $\Vc=-s_z-rs_x$. Its eigenvectors are independent of $m$. Mass modulation preserves the two spin sectors and gives no channel conversion despite finite spin mixing.

\section{Near-total time reflection and microwave emulation}
\label{app:totalreflection}
\label{app:photonic}
\subsection{Control path for near-total reflection}
Antiparallel velocity axes in the parallel frame give $R_{\rm geo}=1$ in Eq.~\eqref{app:eq:geo}.
Set $(\lambda_{\rm SO},\lambda_R)/J=(0.50,0.35)$. A reference ellipse for the valley gap-closing curve is $(m/J)^2+C(g_y/J-g_c)^2=a_m^2$. Here $g_c$ is its hopping-coordinate center, $a_m$ the mass semiaxis, and $a_g=a_m/\sqrt C$ the hopping semiaxis. The coefficients are $C\simeq0.969$, $g_c\simeq-0.450$, and $a_m\simeq2.53$. The interior curve in Fig.~\ref{fig:microwave_controls}(a) with fractional radius $\rho$ and angular coordinate $\theta$ is
\begin{equation}
\begin{aligned}
m/J&=a_m\rho\cos\theta,\\
g_y/J&=g_c+a_g\rho\sin\theta .
\end{aligned}
\label{app:eq:controlcurve}
\end{equation}
The condition that the initial and final velocity axes differ by $\pi$ in the parallel frame fixes $\rho\simeq0.271$. The initial setup has QSH index $\nu=1$, and the minimum direct bulk gap along the path is approximately $1.39J$, preserving the QSH phase throughout the ramp. Let $a_\rho=\arcsin[-g_c/(a_g\rho)]$ and $s=t/T$. The angular coordinate evolves as
\begin{equation}
\theta(t)=\pi-a_\rho+
(2a_\rho-3\pi)(10s^3-15s^4+6s^5).
\label{app:eq:arcclock}
\end{equation}
Both first and second time derivatives of the control waveforms in Fig.~\ref{fig:microwave_controls}(b) vanish at the start and end of the ramp. The controls range over $m/J\in[-0.69,0.69]$ and $g_y/J\in[-1.15,0.25]$. The initial and final masses are $\mp0.52J$ and both initial and final hopping amplitudes have $g_y=0$.

The velocity axis rotates by $\pi$ in the parallel frame, with a crossing-to-bulk gap of about $0.68J$. Figure~\ref{fig:microwave_controls}(c) shows full-ribbon evolution at $q=10^{-6}$ approaching unit reflection as bulk excitation decreases within the interval $\Delta_b^{-1}\ll T\ll\Delta_e(q)^{-1}$. Near-total reflection is thus reached with negligible bulk excitation and the lossless packet in Fig.~\ref{fig:two}(d) has reflection $R\simeq99.9\%$ at $T=80$.

\subsection{Microwave emulation with effective fermionic TRS}
The electronic calculation has fermionic TRS. Helical microwave transport and photonic realizations of effective fermionic TRS motivate emulation with classical modes~\cite{RazoLopez2025,He,Maczewsky}. Here we specify a coupled-mode proposal and simulate a mass-only example with loss and finite control response.

At each lattice site, two microwave modes represent the pseudospin components $\uparrow$ and $\downarrow$. The coherent rotating-frame Hamiltonian obeys effective fermionic TRS with $\widetilde\Theta=(I_{\rm sites}\otimes is_y)K$ and $\widetilde\Theta^2=-I$. A directed hopping block must satisfy $s_yh_{ij}^*s_y=h_{ij}$ and $h_{ji}=h_{ij}^\dagger$, giving
\begin{equation}
h_{ij}=\begin{pmatrix}\alpha&\beta\\-\beta^*&\alpha^*\end{pmatrix}.
\label{app:eq:pairedlinks}
\end{equation}
For a nearest-neighbor unit vector $\bm d=(d_x,d_y)$ from source to destination, $\alpha=-1$ and $\beta=\lambda_R(-d_x+id_y)$. For a next-nearest neighbor, $\alpha=i\lambda_{\rm SO}\nu_{ij}$ and $\beta=0$. The sign $\nu_{ij}=\pm1$ gives the chirality of the next-nearest-neighbor hopping. The intracell $A\to B$ control adds $g_y$ to $\beta$, and the mass gives on-site detunings $+mI_2$ and $-mI_2$ on the two sublattices.

A fixed reciprocal bosonic antiunitary $T_b^2=+1$ imposes an additional constraint if it is preserved together with an effective fermionic antiunitary $\widetilde\Theta^2=-1$. Choose a basis with $T_b=K$ and write $\widetilde\Theta=\mathsf U K$, so $\mathsf U\mathsf U^*=-I$. If both symmetries hold instantaneously, $H(\bm k)^*=H(-\bm k)$ and $\mathsf U H(\bm k)^*\mathsf U^\dagger=H(-\bm k)$ imply $[\mathsf U,H(\bm k)]=0$. Kramers partners belong to opposite eigenvalue sectors of this conserved unitary, which prevents conversion between them under a control preserving both fixed symmetries. The pump-controlled couplers break microscopic bosonic TRS while preserving the effective fermionic TRS of the coherent Hamiltonian.

For laboratory modes with angular frequencies $\omega_a,\omega_b$ and annihilation operators $a_a,a_b$, a modulation
$2g\cos[(\omega_a-\omega_b)t_p-\phi]
(a_a^\dagger a_b+\mathrm{h.c.})$
produces $ge^{i\phi}a_a^\dagger a_b+\mathrm{h.c.}$ in the rotating frame under the rotating-wave approximation. Here $g$ and $\phi$ set the hopping magnitude and phase, and $t_p$ is physical time. Phase-controlled microwave couplings have component-level precedent~\cite{Roushan}. Taking $J_{\rm Hz}=4\,\mathrm{MHz}$ gives couplings and sublattice detunings of a few megahertz and a duration $T=80$ corresponds to $3.2\,\mu\mathrm{s}$.

\subsection{Uniform loss and finite detuning response}
For loss rate $\kappa=\omega_0/Q_L$, the physical-time coupled-mode equation is
\begin{equation}
i\partial_{t_p}\bm a=[2\pi J_{\rm Hz}H(t_p)-i\kappa I/2]\bm a.
\label{eq:microwave}
\end{equation}
The vector $\bm a$ contains rotating-frame mode amplitudes, $\omega_0=2\pi f_0$ with $f_0=5\,\mathrm{GHz}$, and $Q_L=2\times10^5$ is the loaded quality factor. Uniform loss factors out as $\bm a=e^{-\kappa t_p/2}\bm\psi$, where $\bm\psi$ follows the coherent Hamiltonian. For incident norm $N_{\rm in}$ before the ramp, the surviving norm $N_{\rm surv}=\|\bm a\|^2$ satisfies $N_{\rm surv}/N_{\rm in}=e^{-\kappa t_p}$. The dissipative generator itself is not TRS. Conditional mode probabilities follow coherent evolution, while absolute reflected weight obeys $R_{\rm abs}=(N_{\rm surv}/N_{\rm in})R_{\rm cond}$.

Define dimensionless laboratory time $\tau_p=2\pi J_{\rm Hz}t_p$ and response time $\tau_r=2\pi J_{\rm Hz}\tau_c$, with $\tau_c=20\,\mathrm{ns}$. The mass is
$m_{\rm target}(\tau_p)=m_i+(m_f-m_i)f(\tau_p/T_{\rm cmd})$ for $0\le\tau_p\le T_{\rm cmd}=20$, using the same raised-cosine profile.
The filtered mass obeys $\tau_r\partial_{\tau_p}m_{\rm filt}+m_{\rm filt}=m_{\rm target}$, with $m_{\rm filt}(0)=m_i$, and enters $H$. After the ramp, the target is held at $m_f$ and the filtered mass relaxes toward it. The ramp lasts $0.80\,\mu\mathrm{s}$ and readout at $\tau_p=90$ occurs at $3.6\,\mu\mathrm{s}$.

\begin{figure}[t]
\centering
\includegraphics[width=.98\textwidth]{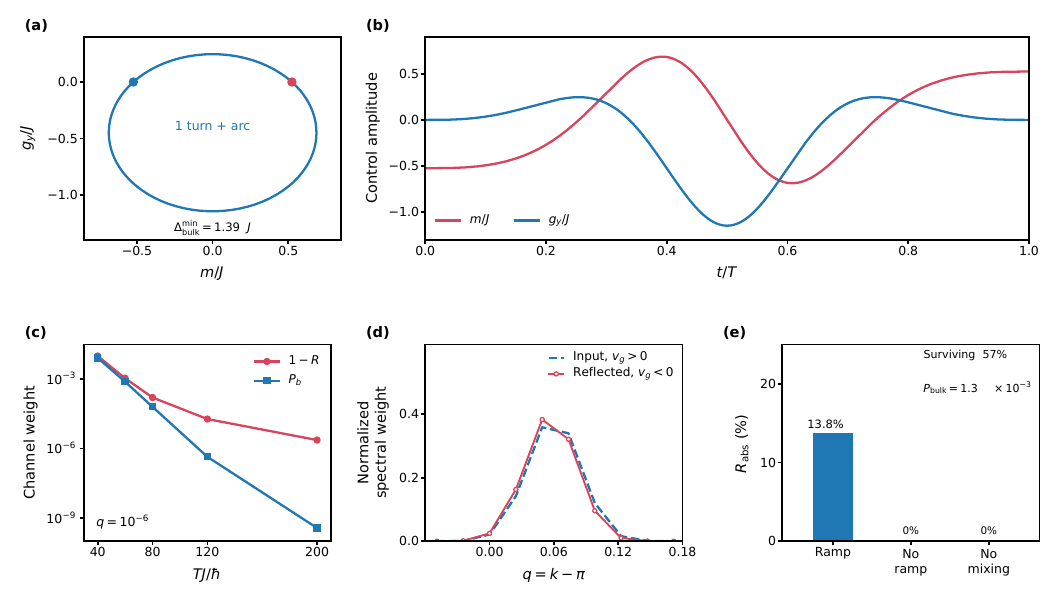}
\caption{Near-total reflection and microwave emulation. \textbf{(a)} Control path for lossless near-total reflection. \textbf{(b)} Control waveforms from Eq.~\eqref{app:eq:arcclock}. \textbf{(c)} Reflection deficit $1-R$ and bulk excitation at $q=10^{-6}$. \textbf{(d)} Separately normalized input and reflected spectra for the lossy mass-only example. \textbf{(e)} Absolute reflected weights for that example with no-ramp and no-mixing.}
\label{fig:microwave_controls}
\end{figure}

The mass-only example changes $m/J$ from approximately $-0.67$ to $0.67$ at $g_y=0$, with $\lambda_{\rm SO}=\lambda_R=0.35J$. It starts in the positive-velocity edge channel. At readout, the surviving norm is about $0.57$, giving $R_{\rm abs}\simeq13.8\%$ and $R_{\rm cond}\simeq24.2\%$. The bulk and opposite-edge weights relative to $N_{\rm in}$ are negligible. Figure~\ref{fig:microwave_controls}(d) shows that the reflected spectrum remains in the incident momentum window. Figure~\ref{fig:microwave_controls}(e) compares the absolute reflected weight with no-ramp and no-mixing controls, which suppress conversion. The zero-Rashba control starts in the positive-velocity eigenmode of its own Hamiltonian with the same envelope.

\end{document}